\documentclass[preprint,aps,amssymb,showpacs,superscriptaddress,nofootinbib]{revtex4}

\usepackage{graphics}
 
\usepackage{makeidx}
\usepackage{epsfig}
\usepackage{subfig}
\usepackage{colortbl}
\usepackage{colordvi}
\usepackage{verbatim}
\usepackage{amsmath, amsthm}
\usepackage{enumerate}
\usepackage{wrapfig}
\usepackage{hyperref}
\usepackage{dcolumn}
\usepackage{bm}
\usepackage{bbm}
\usepackage{amssymb,mathrsfs}

\def\k{\mathbf{k}}

\def\x{\mathbf{x}}

\begin{document}
 
\title{Polymer Quantization predicts radiation in inertial frames}


\author{Nirmalya Kajuri}
\email{nirmalya@imsc.res.in}
 \affiliation{The Institute of Mathematical Sciences\\
CIT Campus, Chennai-600 113, INDIA.}

\date{\today}

\begin{abstract} We investigate the response of an Unruh-DeWitt detector coupled to a polymer quantized massless scalar field in flat spacetime, using the propagator obtained by Hossain, Husain and Seahra. As this propagator violates Lorentz invariance, frames moving at different constant velocities are no longer equivalent. This means that it is possible in principle for even an observer moving at constant velocity to detect radiation. We show that such an observer indeed detects radiation. Remarkably, we show that the rate of this radiation does not decrease with the decrease in the characteristic length scale of polymer quantization. Thus the radiation cannot be suppressed by making the polymer length scale arbitrarily small. Our results should bring this theory within the ambit of low-energy experiments and place a lower limit on the characteristic polymer length scale.

\end{abstract}
 
\maketitle
\section{Introduction}

In \cite{ Hossain:2010eb} Hossain, Husain and Seahra introduced a novel quantization for scalar field theories. This quantization replicates some features of the quantization in Loop Quantum Gravity and has been referred to as 'polymer quantization' of the scalar field. For now we will call this the 'momentum space polymer quantization' to distinguish it from a different 'position space' quantization for scalar fields  that is also referred to as 'polymer quantization' in the LQG literature. 

In \cite{ Hossain:2010eb} the quantization was carried out on a massless scalar field in Minkowski spacetime. First, one Fourier decomposed a massless Klein Gordon field on a Minkowski background. This gives a system of uncoupled simple harmonic oscillators, one at each point of momentum space. Each of these oscillators was then quantized using the \textit{polymer particle representation}, a LQG-like quantization for non-relativistic particles introduced in \cite{ Ashtekar:2002sn}. This quantization involves the introduction of a length scale $\lambda_*$ to define certain observables. The limit  $\lambda_* \rightarrow 0$ corresponds to the usual, Schrodinger quantized harmonic oscillator \footnote{ This limit cannot actually be taken in the polymer Hilbert Space, as we will show in the next section. However $\lambda_*$  can be made smaller and smaller arbitrarily, resulting in better and better agreement with the standard results.}. The propagator for the scalar field was obtained and was seen to violate Lorentz invariance.

Recently in \cite{Hossain:2014fma}, it was claimed that the Unruh Effect vanishes in this momentum space polymer field theory. However the result was questioned in \cite{Rovelli:2014gva}. In \cite{Hossain:2014fma}  the method of Bogoliubov transformation had been used to probe the existence of Unruh Effect. An alternate approach is to study the response of a detector(usually called the Unruh Dewitt detector) \cite{Unruh:1976db, DeWitt:2003pm} moving in different trajectories in spacetime. We take up this approach in this paper. 

However, we study not only the detectors moving in accelerated trajectories but also those moving with constant velocity. As Lorentz invariance is violated it follows that the principle of relativity does not apply for this theory. As we'll see, there is a preferred frame chosen in the process of quantization. All frames, even those moving with constant velocity with respect to the preferred frame, are inequivalent. Different inertial observers will disagree on the vacuum. Therefore it is quite possible that an observer moving with constant velocity with respect to the preferred frame will also observe a phenomenon analogous to the Unruh Effect.

 We study the response of the Unruh Dewitt detector in three different frames - (i) The detector is at rest in the preferred frame (ii) The detector is moving with constant speed with respect to the preferred frame and (iii) the detector is moving with constant acceleration with respect to the preferred frame. 

 We show that, just as in the case of the Fock quantized scalar field theory, the detector in (i) does not click \footnote{By 'click' we mean 'make a transition from a lower to higher energy level'.} while the accelerated detector of (iii) does. A more interesting result is found for (ii) where we show that even an 'inertial'\footnote{We'll use the term 'inertial frame' to simply mean the frame of a constant velocity observer i.e the frame of an observer moving along one of the geodesics of Minkowski spacetime.} detector will click while moving through a vacuum.

Even more remarkably, we show that this rate of clicking cannot be made smaller by making the polymer scale $\lambda_*$ arbitrarily small- instead it increases with the decrease in $\lambda_*$. \textit{Thus this theory disagrees with the usual results of the Fock quantized theory in the domain of validity of the latter} \footnote{This is all the more remarkable because the polymer propagator was shown to agree with the ususal Feynman propagator in this domain \cite{ Hossain:2010eb}.}. Therefore our work should bring this theory within the ambit of \textit{low energy} experiments.  We expect our results to place strong constraints on the \textit{lower limit} of $\lambda_*$. 

(\textit{Note Added:} Progress along this line has already been made. In a paper that appeared after our submission\cite{Husain:2015tna}, a more detailed investigation of transition rates for the inertial Unruh-DeWitt detector was undertaken, both analytically and numerically. This paper bears out our result that the inertial detector coupled to a polymer quantized scalar field can click. Furthermore, it showed that there exists critical velocity $\beta_c =1.3675$ for detectors. A detector moving below this velocity (with respect to the preferred frame) will not get spontaneously excited. However, a detector moving with a speed above $\beta_c$ will click, even when the detector's energy gap is very small or the polymer length scale $\lambda_*$is very small. In fact the rate of transitions for such a detector was shown to be proportional to $\frac{1}{\lambda_*}$. It is to be noted that the critical velocity discovered in \cite{Husain:2015tna} is well within the range of present day experiments.)

Before going into the details, we should clarify the implications of our results for Loop Quantum Gravity. There are none. We always work in a Minkowski background and gravity does not enter our calculations. However, there is a different way Loop Quantum Gravity may enter the discussion. The usual Fock quantization method requires a background geometry to be present. Information about this geometry enters the construction of the field theory Hilbert Space. In LQG however the background geometry is itself quantized. Therefore, when quantizing a coupled matter-gravity system, one cannot use Fock quantization for matter fields. One must use a 'background independent' quantization for matter fields. The momentum space polymer quantization was introduced in \cite{ Hossain:2010eb} as a quantization for scalar field theory that is compatible with LQG.  However the construction is not entirely background independent as the mode decomposition depends on the background. There exists a different quantization for scalar fields, which is entirely background independent and perfectly compatible with LQG. This is also called  polymer quantization in the literature. This 'position space polymer quantization' for scalar fields was introduced in \cite{Ashtekar:2002vh}. Here one directly constructs a Hilbert Space as a space of functionals on scalar fields  equipped with a diffeomorphism invariant inner product. As the Hilbert Space construction is background independent, the only way information about the background enters this theory is through the Hamiltonian. Defining this Hamiltonian involves the introduction of a scale in this case as well. 

The 'momentum space' and 'position space' polymer quantizations can be seen to be dynamically different. One may see this by directly comparing the Hamiltonians obtained by carrying our both the quantizations in a Minkowski spacetime. Interestingly, in both cases one obtains Lorentz violation \cite{Kajuri:2014kva}. The point to take away is that Loop Quantum Gravity does not uniquely single out the momentum space polymer quantization as the appropriate quantization for scalar fields. Conversely, any result about this quantization does not necessarily have any bearing on LQG. From hereon, we will drop the prefix 'momentum space' and simply use the term 'polymer quantization' to describe the quantization of \cite{ Hossain:2010eb}.

With this caveat out of the way, we are now ready to present the details of our work. The next section introduces the polymer propagator obtained by Hossain, Husain and Seahra. In section (3) we briefly recall the analysis of the Unruh-Dewitt detector coupled to a Fock quantized scalar field. Section (4) presents our analysis of the Unruh Dewitt detector coupled to a Polymer scalar field. We summarize our results and present our conclusions in section (5). We will take the space-time signature to be (-+++).

\section{The Polymer Propagator}

\subsection*{A. Polymer Quantized Harmonic Oscillator}
The distinguishing feature of the polymer particle representation is that both position and momentum operators cannot be well-defined on the particle Hilbert Space. One may choose one of them to be well-defined. If position is chosen to be well defined, momentum will not be well-defined. Instead the family of translation operators will be well defined. As we show below, an \textit{approximate} momentum operator may be defined from these translation operators by introducing some scale $\lambda_*$. The use of this approximate operator in the Hamiltonian leads to a modification of the energy spectrum.

Let us describe the construction in more details. To construct the Hilbert Space, we first choose a countable set, $\gamma = \{ {x}_j, x_j \epsilon \mathbb{R} \}$ and define a set Cyl$_{\gamma}$ of
linear combinations of the form: Cyl$_{\gamma} := \{\sum_{j} f_j e^{i  {x}_j {p}}, f_j \in \mathbb{C}  \}$. Then we define the set of functions of $ p$, Cyl := $\cup_{\gamma}$Cyl$_{\gamma}$. The inner product on this set is chosen to be
\begin{align} 
(e^{i{x_i} {p}}, e^{i {x_j} p}) = \delta_{ {x_i},{x_j}} 
\end{align}
  $\{e^{ixp}~/x \in \mathbb{R} \}$  form an uncountable basis of this space and we denote them as the kets $|x\rangle$. The completion of Cyl w.r.t this inner product is our requisite Hilbert Space  $H_{\mathrm{poly}}$: $\overline{\mathrm{Cyl}} =: H_{\mathrm{poly}}$. 

On this Hilbert Space we have the basic operators:
\begin{align} 
\hat{x}|x\rangle = x|x\rangle
\end{align}
and
\begin{align}  \hat{V}(\lambda)|x\rangle = |x -\lambda\hbar\rangle \end{align}

As $ \hat{V}(\lambda)$ is not weakly continuous in $\lambda $ a momentum operator cannot be defined. We can however define an approximate momentum operator by choosing some scale $\lambda_*  $:  \begin{align}\hat{p}|_{\lambda_*  }=\frac{\hat{V}(\lambda_* )-\hat{V}(-\lambda_* )}{2\lambda_* i}\end{align}  
Here we work with natural units and $\lambda_*  $ has the dimensions of length.

We now consider the case of the simple harmonic oscillator. The simple harmonic oscillator Hamiltonian is defined using the approximate momentum operator given above and reads:  
\begin{align}
\hat{H} = \frac{1}{8m\lambda_*^2}(2 -\hat{V}(\lambda_* )-\hat{V}(-\lambda_* )) + \frac{m\omega^2 x^2}{2}
\end{align}

The time independent Schrodinger equation is modified to: 
\begin{align}
 \frac{1}{8m\lambda_*^2}( 2- 2\cos(2\lambda_*p))\psi - \frac{m\omega^2 }{2}\frac{\partial^2 \psi}{\partial p^2} = E\psi
\end{align}
 
This can be transformed into the Mathieu equation through the following redefinitions: 
\begin{align}
u = \lambda_*p +\pi/2  \,, \qquad \alpha = 2E/{g\omega} - 1/{2g^2} \,, \qquad g =m\omega \lambda_*^2 
\end{align}

With these redefinitions the above equation takes the standard form of the Mahtieu equation:
\begin{align}
\psi^{\prime \prime}(u) + (\alpha - \frac{1}{2}g^{-2}\cos(2u))\psi(u) = 0
\end{align}

This equation admits periodic solutions for certain values of $\alpha$:
\begin{align}
\psi_{2n}(u) &= \pi^{-1/2} ce_n(1/{4g^2}, u),\qquad \alpha =A_n(g)\\
\psi_{2n+1}(u) &= \pi^{-1/2} se_{n+1}(1/{4g^2}, u),\qquad \alpha =B_n(g)
\end{align}
where $ ce_n, se_n $ are respectively the elliptic cosine and sine functions and $A_n, B_n$ are the Matheiu characteristic value functions.
Now we may express the energy eigenvalues of the polymer harmonic oscillator: 
\begin{align}
\frac{E_{2n}}{\omega}=\frac{2g^2A_n(g)+1}{4g}
\end{align}
\begin{align}
\frac{E_{2n+1}}{\omega}=\frac{2g^2B_{n+1}(g)+1}{4g}
\end{align}

Analytic approximations are available for these functions for the asymptotic cases $g<<1$ and $g>>1$ \cite{ Hossain:2010eb}. For further information on the physics of the polymer quantized harmonic oscillator we direct readers to \cite{ Ashtekar:2002sn, G.:2013lia}.

\subsection*{B. Polymer Propagator for Scalar Field}

Now we turn to canonically polymer quantizing a massless scalar field. This had been introduced in \cite{ Hossain:2010eb}, whose treatment we now follow. The first step here is to choose a slicing of spacetime. We choose one where the 3-space is flat. This corresponds to choosing some global inertial co-ordinate system. 
\begin{align}
\label{hamiltonian} H = \int d^3x \left( \frac{\pi^2}{2} + \frac{(\nabla \phi)^2}{2} \right) 
\end{align}

Now one proceeds to quantizing the theory in this frame. If the quantum theory is Lorentz invariant then this choice of an inertial frame would be inconsequential - the quantum theory would be the same no matter which inertial frame one chooses to quantize in. However the propagator from polymer quantization will violate Lorentz invariance. This means that the resulting quantum theory will \textit{depend} upon the frame chosen for quantization. We will call this the preferred frame.

To polymer quantize the above Hamiltonian, we first Fourier expand it. This gives us a system of uncoupled Harmonic oscillator Hamiltonians: 
\begin{align}
H_\k =\frac{\pi_\k^2}{2} +\frac{|\k|^2\phi_\k^2}{2}
\end{align}
Now one polymer quantizes each oscillator. Just as $\hat{x},\hat{V}$ were the basic operators in the previous case, we will have $\hat{\phi}_\k,\hat{U}_\k(\lambda)$ as the basic operators here. Here $\hat{U}_\k(\lambda) = e^{i\lambda\pi_\k}$ classically. As earlier one can define an approximate momentum operator by introducing a scale $\lambda_*$. Note that in this case $\lambda_*$ has the dimension of $\sqrt{\text{length}}$ . This gives a polymer quantum Hamiltonian $\hat{H_\k}$ which can be mapped to the polymer SHO Hamiltonian of the previous section by putting $m$=1 and identifying $|\k|=\omega$. So each $\hat{H_\k}$ will have the same spectrum as the oscillator of the last section with: 
$$ g= \lambda_*^2|\k|=\frac{|\k|}{M_*} = \frac{\text{frequency}}{\text{polymer mass scale}} $$

where $M_*= \lambda_*^{-2}$ is termed as the polymer mass scale and it's inverse may be called the polymer length scale. So when $g$ is small it means that the frequencies are small compared to the polymer mass scale and we should expect the polymer theory to reproduce the results of the usual Fock quantized field theory in this regime. 

 The polymer vacuum is the state where all the oscillators are at ground state: 
$$ |0\rangle = \prod_k |0_{\k}\rangle$$
We may obtain the two point function $\langle 0|\hat{\Phi}(t,\vec{x}) \hat{\Phi}(t',\vec{x'})|0\rangle$ from the modified spectrum.
We start by writing\footnote{ A careful derivation from step \eqref{hamiltonian} to here would involve first introducing a fiducial volume V in the definition of the Fourier transform, which can eventually be taken to be infinity \cite{Hossain:2015xqa}.}
 
$$\langle 0|\hat{\Phi}(t,\x) \hat{\Phi}(t',\x')|0\rangle = \int \frac{d^3\k}{(2\pi)^3} ~  D_{\k}(t,t') e^{i {\k}\cdot(\x-\x')}$$

Where

$$ D_{\k}(t,t') = \langle 0_{\k}| e^{i\hat{H}_{\k}t} \hat{\phi}_{\k} e^{-i\hat{H}_{\k}t}e^{i\hat{H}_{\k}t'} \hat{\phi}_{\k} e^{-i\hat{H}_{\k}t'}|0_{\k}\rangle$$

 Using the eigenspectrum of the Hamiltonian and expanding the state $\hat{\phi}_{\k} |0_{\k}\rangle$ 
in the basis of energy eigenstates as 
$\hat{\phi}_{\k}|0_{\k}\rangle = \sum_{n} c_n |n_{\k}\rangle$, 
 we can evaluate the above equation to obtain
\begin{equation}
\label{DkFunctionGeneral}
D_{\k}(t-t') \equiv D_{\k}(t,t') = \sum_{n} |c_n|^2 e^{-i\Delta E_n (t-t')},
\end{equation}
where $\Delta E_n \equiv E_n^{(\k)} - E_0^{(\k)}$ and
$c_n = \langle n_{\k}| \hat{\phi}_{\k} |0_{\k}\rangle$.

Thus we have the following expression for the two point function: 
\begin{align}
\label{twopoint}
\langle 0|\hat{\Phi}(t,\x) \hat{\Phi}(t',\x')|0\rangle = \sum_{n} \int \frac{d^3\k}{(2\pi)^3}e^{i {\k}\cdot(\x-\x')}|c_n|^2 e^{-i\Delta E_n (t-t')}
\end{align}

In \cite{ Hossain:2010eb} it was shown that the only non-zero values of $c_n$ are for $c_{4n+3}$ (for $n =0,1,2,...$). 

We note that the definition of the two point function above does \textit{not} involve time ordering. The corresponding expression for a Fock quantized massless scalar field reads:
\begin{align}
\label{focktwopoint}
\langle 0|\hat{\Phi}(t,\x) \hat{\Phi}(t',\x')|0\rangle =   \int \frac{d^3\k}{(2\pi)^3 2|\k|}e^{i {\k}\cdot(\x-\x')}  e^{-i |\k|(t-t')}
\end{align}


\section{The Unruh Dewitt detector}
Let us briefly recall the study of the response of an Unruh Dewitt detector coupled with a Fock quantized massless scalar field.

We consider a point-like detector moving through spacetime along a worldline $x^{\mu}(\tau)$ where $\tau$ is the proper time along its world line. The Hamiltonian of the coupled field-detector system is:

$$ H = {(H_0)}_{\text{detector}} +{(H_0)}_{\text{field}} +\alpha N \phi(x(\tau))$$

where $N$ is a perturbation of ${(H_0)}_{\text{detector}}$ and $\alpha$ a small coupling constant \footnote{In general, the perturbation $N$ should include a switching function $\chi(\tau)$ and the transition rate can be extracted in the limit of adiabatic switching - see for instance \cite{Satz:2006kb, Louko:2007mu}. The authors of \cite{Husain:2015tna} have verified that for appropriate switching functions, the result of this procedure agrees with that obtained by using constant $N$ as we do here.}. The detector may undergo a transition from one energy state to another through its interaction with the field. The probability of observing a transition in the detector from a state of energy $E$ to a state of energy $E+\omega$ is given up to first order in perturbation theory according to:

$$ \text{Prob}(E \rightarrow E+\omega) = \alpha^2|\langle E+\omega| N| E\rangle|^2 \int d\tau d\tau' e^{-i\omega(\tau-\tau')}\langle i|\phi(x(\tau))\phi(x(\tau'))|i\rangle $$

Where $|i\rangle$ denotes the initial state the scalar field was in. The above formula is obtained by first using first order perturbation theory to calculate the probability of a transition where both the detector and the field undergo transitions and then summing over all possible final states for the field. Now we assume the initial state for the field $|i\rangle $ was actually the vacuum state $|0 \rangle$. Then one can use the translational invariance of the vacuum to the re-write the above equation as
\begin{align}
\label{common}
\text{Prob}(E \rightarrow E+\omega) = \alpha^2\langle N\rangle^2 \int d\tau d\tau' e^{-i\omega(\tau-\tau')}\langle 0|\phi(x(\tau) -x(\tau'))\phi(0)|0\rangle 
\end{align}
 Note that till this point we have not made any assumption either about the quantization of the field - it can be either polymer or Fock. We've only assumed translational invariance of the vacuum which holds in both cases. Nor have we assumed anything about the state of motion of the detector - it may be inertial or accelerated. We now assume that the detector is moving in an inertial frame with $ x^\mu(\tau)-x^\mu(\tau') = u^\mu(\tau-\tau')$ with constant $u^\mu$. Then one of the integrals in \eqref{common} becomes trivial and we have:
\begin{align}
\label{common2}
\text{Rate}=\frac{\text{Prob}(E \rightarrow E+\omega)}{\text{proper time}} =\alpha^2{\langle N\rangle}^2 \int d\tau e^{-i\omega\tau}\langle 0|\phi(x(\tau))\phi(0)|0\rangle 
\end{align}

To evaluate this integral it is simplest to go to the rest frame of the detector and use \eqref{focktwopoint}:
\begin{align}
\nonumber \int d\tau e^{-i\omega\tau}\langle 0|\phi(x(\tau))\phi(0)|0\rangle &= \int d\tau e^{-i\omega\tau} \int \frac{d^3\k}{(2\pi)^3 2|\k|}e^{-i {|\k|}\tau}\\
\nonumber &=\int \frac{d^3\k}{(2\pi)^3 2|\k|} 2\pi\delta \left(|\k|+\omega \right)\\
&= \int \frac{|\k|^2}{2|\k|}\delta \left(|\k|+\omega \right)
\end{align}
Where in the second step we performed the integral over $\tau$. As $|\k|$ is always positive the integral vanishes for $\omega > 0$. For $\omega$ less than 0 the integral can be seen to give $-\frac{\omega}{2\pi}$.

So we see that a detector in the rest frame will not click. Because of Lorentz invariance of the Fock quantized field theory, this result obviously extends to all inertial frames. Nevertheless we explicitly demonstrate it here, as this is the result which gets be modified when a polymer quantized theory is considered. We consider a constant velocity frame where the detector moves along the $x^1$-axis. In this frame we have $ t = x^0 =u^0\tau$ and $x^1 = u^1\tau$. Then the above integral becomes:
\begin{align}
\nonumber &\int \frac{d^3\k}{(2\pi)^3 2|\k|} \int d\tau e^{-i\omega\tau} e^{-i|\k|u^0\tau + ik^1u^1\tau}\\
&= \int \frac{d^3\k}{(2\pi)^2 2|\k|}\delta \left(|\k|u^0  - k^1u^1 + \omega \right)
\end{align}

Now as $|\k| \geq k^1 $ by definition and $u^0 > u^1$ for all time-like trajectories this ensures that the integral once again vanishes for $\omega>0$. 

We come to the case of a detector moving with constant acceleration. We can take \eqref{common} as the starting point. One notes that (i) Lorentz boosts generate translations along constantly accelerated worldlines (ii)the  Fock vacuum is invariant under Lorentz boosts. These two facts can be used to express the Prob($E\rightarrow E+ \omega$) as a function of the difference $(\tau'-\tau)$ only. This makes one of the integrals trivial and ensures that the rate of transitions is again given by the formula \eqref{common2}. The only difference now is that the functional dependence of $x^\mu$ on $\tau$ has changed. Specifically, for a detector moving with constant acceleration $a$ along the $x^1$ direction we have 
\begin{align}
\nonumber &t = x^0(\tau)=\frac{1}{a}\sinh(a\tau)\\
&x^1(\tau)=\frac{1}{a}\cosh(a\tau)
\end{align}

Substituting this in \eqref{common2} one may evaluate the rate of transitions for the accelerated detector. This turns out to be: 
\begin{align}
\nonumber \text{Rate} = &\alpha^2 \langle N\rangle^2 \frac{\omega}{2\pi} \frac{e^{\frac{2\pi|\omega|}{a}}} {e^{\frac{2\pi|\omega|}{a}}-1}, \qquad \omega < 0 \\
=&\alpha^2 \langle N\rangle^2 \frac{\omega}{2\pi} \frac{ 1} {e^{\frac{2\pi|\omega|}{a}}-1}, \qquad \omega > 0
\end{align}

So we see that the accelerated detector clicks. It can be shown that at equilibrium the probability of occupancy of the states of the detector will be given by the Boltzmann distribution, with temperature 
 $\frac{a}{2\pi}$. Thus the accelerated observer will find itself in an environment equivalent to a \textit{thermal} bath. 

To summarize, we studied the response of the Unruh Dewitt detector coupled to a Fock quantized scalar field in three different states of motion - rest, constant velocity and constant acceleration. We saw that in the first two cases (which are of course equivalent here due to Lorentz invariance) the detector does not click while for the final case it does click. Moreover the detector will equilibriate when it's energy states are distributed according to the Boltzmann distribution. Now we will study the same cases for a Unruh Dewitt detector coupled to a polymer quantized scalar field.

\section{Unruh Dewitt detector coupled to a polymer Scalar Field}
As we have seen before, Lorentz invariance is absent in the polymer quantized scalar field theory. Thus we must consider three different cases here: (i) the detector is at rest in the preferred frame (ii) the detector is moving with constant acceleration with respect to the preferred frame and (iii)the detector is moving with constant velocity with respect to the preferred frame. Unlike the Lorentz invariant Fock quantized theory, cases (i) and (iii) are inequivalent for the polymer quantized theory. Let us repeat the analysis outlined in the last section to each of these cases. As we are interested in the question of whether the detector clicks or not we will restrict the analysis to the $\omega <0$ case. 

\subsection*{Detector at rest}
The analysis for an inertial detector (i.e a detector at rest in the preferred frame or moving with constant velocity) is identical to that for the Fock case up to \eqref{common2}:
\begin{align}
\nonumber
\text{Rate}=\frac{\text{Prob}(E \rightarrow E+\omega)}{\text{proper time}} =\alpha^2{\langle N\rangle}^2 \int d\tau e^{-i\omega\tau}\langle 0|\phi(x(\tau))\phi(0)|0\rangle 
\end{align}
Only instead of the Fock two point function one must now use the polymer two point function given by \eqref{twopoint}. This gives:

\begin{align}
\label{common3}
\alpha^2 \langle N\rangle^2 \sum_{n}\int d\tau e^{-i\omega\tau} \int \frac{d^3\k}{(2\pi)^3}e^{i {\k}\cdot\x(\tau)}|c_n|^2 e^{-i\Delta E_n t(\tau)}
\end{align}

In the rest frame $\x =0, t=\tau$ and we have 
\begin{align}
\nonumber
&\alpha^2 \langle N\rangle^2 \sum_{n}\int \frac{d^3\k}{(2\pi)^3}|c_n|^2 \int d\tau e^{-i\omega\tau} e^{-i\Delta E_n \tau}\\
&=\alpha^2 \langle N\rangle^2 \sum_{n}\int \frac{d^3\k}{(2\pi)^2}|c_n|^2 \delta \left(\omega + \Delta E_n \right)
\end{align}

As $\Delta E_n $ is always positive we see that the detector does not click in this case. This is expected as the scalar field vacuum was defined in this frame as the state in which all the harmonic oscillators the field was decomposed into are in their ground states. Therefore there can be no transfer of energy from the scalar field to the detector if the field is in this state. 

\subsection*{Detector moving with constant acceleration}
We now come to the case of a detector moving with constant acceleration with respect to the preferred frame. Our aim in this section is to show that the detector clicks or in other words the probability that the detector makes a transition from a lower to higher energy state does not vanish for all accelerated frames. The analysis of the previous section goes through till \eqref{common}

\begin{align}
\nonumber
\text{Prob}(E \rightarrow E+\omega) = \alpha^2\langle N\rangle^2 \int d\tau d\tau' e^{-i\omega(\tau-\tau')}\langle 0|\phi(x(\tau) -x(\tau'))\phi(0)|0\rangle 
\end{align}

The argument that we used in the previous section to derive \eqref{common2} from \eqref{common} relied on the Lorentz invariance of the Fock. Since we no longer have that in the polymer case we must work with \eqref{common}.

From \eqref{twopoint}, \eqref{common} we have 
\begin{align}
\label{accel}
\text{Prob}(E \rightarrow E+\omega) = \alpha^2\langle N\rangle^2\sum_{n}\frac{d^3\k}{(2\pi)^2}|c_n|^2 \int d\tau d\tau' e^{-i\omega(\tau-\tau')}e^{i {\k}\cdot(\x-\x')}|c_n|^2 e^{-i\Delta E_n (t-t')}
\end{align}

As in the previous section we will consider a detector moving with constant acceleration $a$ along the $x^1$ direction where we have 
\begin{align}
\nonumber t = x^0(\tau)&=\frac{1}{a}\sinh(a\tau)\\
\nonumber  x^1(\tau)&=\frac{1}{a}\cosh(a\tau)
\end{align}
 Substituting this in \eqref{accel} we find that the integrals over $\tau, \tau'$ read:
\begin{align}
\int d\tau d\tau' e^{-i\omega(\tau-\tau')}e^{i  k^1(\frac{1}{a}\cosh(a\tau)-\frac{1}{a}\cosh(a\tau') )}  e^{-i\Delta E_n (\frac{1}{a}(\sinh(a\tau)- \sinh(a\tau'))}
\end{align}
We may write the above as 
\begin{align}
&\nonumber \int d\tau e^{if(\tau)}\int d\tau' e^{-if(\tau)'} \\
&=\left(\int d\tau \cos (f(\tau)) \right)^2 +\left(\int d\tau  \sin (f(\tau)) \right)^2
\end{align}

Where $$f(\tau) = -\omega\tau + \frac{k^1}{a}\cosh(a\tau)-\frac{\Delta E_n(\lambda_*^2|\k|)}{a}\sinh(a\tau)$$ 

Thus \eqref{accel} now reads : 
\begin{align}
\label{accel2}
\text{Prob}(E \rightarrow E+\omega) = \alpha^2\langle N\rangle^2\sum_{n}\frac{d^3\k}{(2\pi)^2}|c_n|^2  \left(\int \cos (f(\tau)) \right)^2 +\left(\int \sin (f(\tau)) \right)^2
\end{align}

We find that the integrand of the $\k$ integral is always positive. Thus we have established quite generally that the only way the integral can vanish is if $\int d\tau\cos (f(\tau))$ and $\int d\tau\sin (f(\tau)$ both vanish for all values of $\k$ and $n$. The condition for complete vanishing of Unruh Effect is thus  $$\int d\tau\cos\left(-\omega\tau + \frac{k^1}{a}\sinh(a\tau)-\frac{\Delta E_n}{a}\cosh(a\tau)\right) =0 $$
and
$$ \int d\tau\sin \left(-\omega\tau + \frac{k^1}{a}\cosh(a\tau)-\frac{\Delta E_n}{a}\sinh(a\tau)\right) = 0$$

for all values of $k^1,k^2,k^3,n,a, \lambda_*$ and all $\omega>0$. That this is not true can be easily checked numerically. 

We have thus established that the detector will in general click for an accelerated detector coupled to a polymer scalar field. We did not however check whether the detector in this case experiences a thermal bath or not. This would be the case when the probabilities of the detector occupying the different energy states is given by the Boltzmann distribution. It is quite possible that this no longer holds in the polymer case. Indeed the detector experiencing a thermal bath is equivalent to the analytic continuation of the two point function satisfying the KMS condition and it has been argued in \cite{Hossain:2015xqa} that the KMS condition fails for the polymer quantized theory. 


\subsection*{Detector moving with constant velocity}

Finally we come to the case of the detector moving with constant velocity with respect to the preferred frame. We can start our analysis from \eqref{common3}

\begin{align}
\nonumber
\alpha^2 \langle N\rangle^2 \sum_{n}\int d\tau e^{-i\omega\tau} \int \frac{d^3\k}{(2\pi)^3}e^{i {\k}\cdot\x(\tau)}|c_n|^2 e^{-i\Delta E_n t(\tau)}
\end{align}

Let us the constant velocity frame along the x-axis. We have $ t = x^0 =u^0\tau$ and $x^1 = u^1\tau$ where $u^0, u^1$ are constant.

Substituting this in \eqref{common3} we have 

\begin{align}
\nonumber
&\alpha^2 \langle N\rangle^2 \sum_{n} \int \frac{d^3\k}{(2\pi)^3}|c_n|^2 \int d\tau e^{-i\omega\tau}e^{i k^1u^1\tau }e^{-i\Delta E_n u^0 \tau }\\
&=\alpha^2 \langle N\rangle^2 \sum_{n} \int \frac{d^3\k}{(2\pi)^2}|c_n|^2 \delta \left(\Delta E_n u^0 -k^1u^1 + \omega \right)
\end{align}

Thus whether the detector clicks or not comes down to whether the term inside the delta function can be zero for $\omega>0$. That is the condition for the detector to click is that the following inequality is satisfied for some combination of the parameter values.
\begin{align}
\Delta E_n(\lambda_*^2|\k|) u^0 -k^1u^1 <0  
\end{align}

We may re-write this as 
\begin{align}
 \label{ether}
g_n(|\k|, \beta, \eta, \lambda_*) = \frac{1}{1-\beta^2} (\Delta E_n(\lambda_*^2|\k|) - \eta\beta|\k| ) <0 
\end{align}

where $\beta = u^1/u^0 \leq 1$ and $\eta= \frac{k^1}{|\k|} = \sin\theta\cos\phi \leq 1$ $\theta, \phi$ being the polar and azimuth angles in the momentum space respectively. Thus we now need to see if there is some $(|\k|, \beta, \eta, \lambda_*, n)$ for which this inequality holds. However we have seen before the non-zero values of the coefficient $c_n$ are for $c_{4m+3}$ (for $m =0,1,2,...$). So we need to only check this inequality for the functions $g_{4m+3}$.

Plotting $g_{4m+3}$ with $|\k| $ for different values of $m$ we find that for all $(\beta, \eta, \lambda_*)$ the function $g_m$ is always positive except when m=0. $g_3$ does satisfy the inequality \eqref{ether}. Moreover this property of $g_3$ is robust under changes in $\beta \hphantom{a} \text{and} \lambda_*$ as we show in the figures. Thus a detector moving in a constant velocity frame will in general click, if the field it couples to is a polymer quantized scalar field.  

From the figures it is also clear that there is an upper bound on the maximum negative value of $\omega$ given by the minima of $g_3$, which we will call $\omega_*$. There will therefore be no transition from a lower to a higher energy state that involves energy exchanges greater than $|\omega_*|$. This means that if the minimum energy gap between any two states of a detector is less than $|\omega_*|$ then it will not click. We see from the plots however that $|\omega_*|$ increases with the decrease in the  polymer length scale  $\lambda_*^2$. 

We also see from the figure that the number of modes contributing to the transitions decreases with the increase in $\lambda_*$ (decrease in  $M_*$). This suggests that such transitions would be easier to detect with the \textit{decrease} in  $\lambda_*$. Experiments that show the absence of clicking should therefore put a \textit{lower bound} on  $\lambda_*$. 
\begin{figure}[]
\centering
{}{\includegraphics[width=0.45\textwidth]{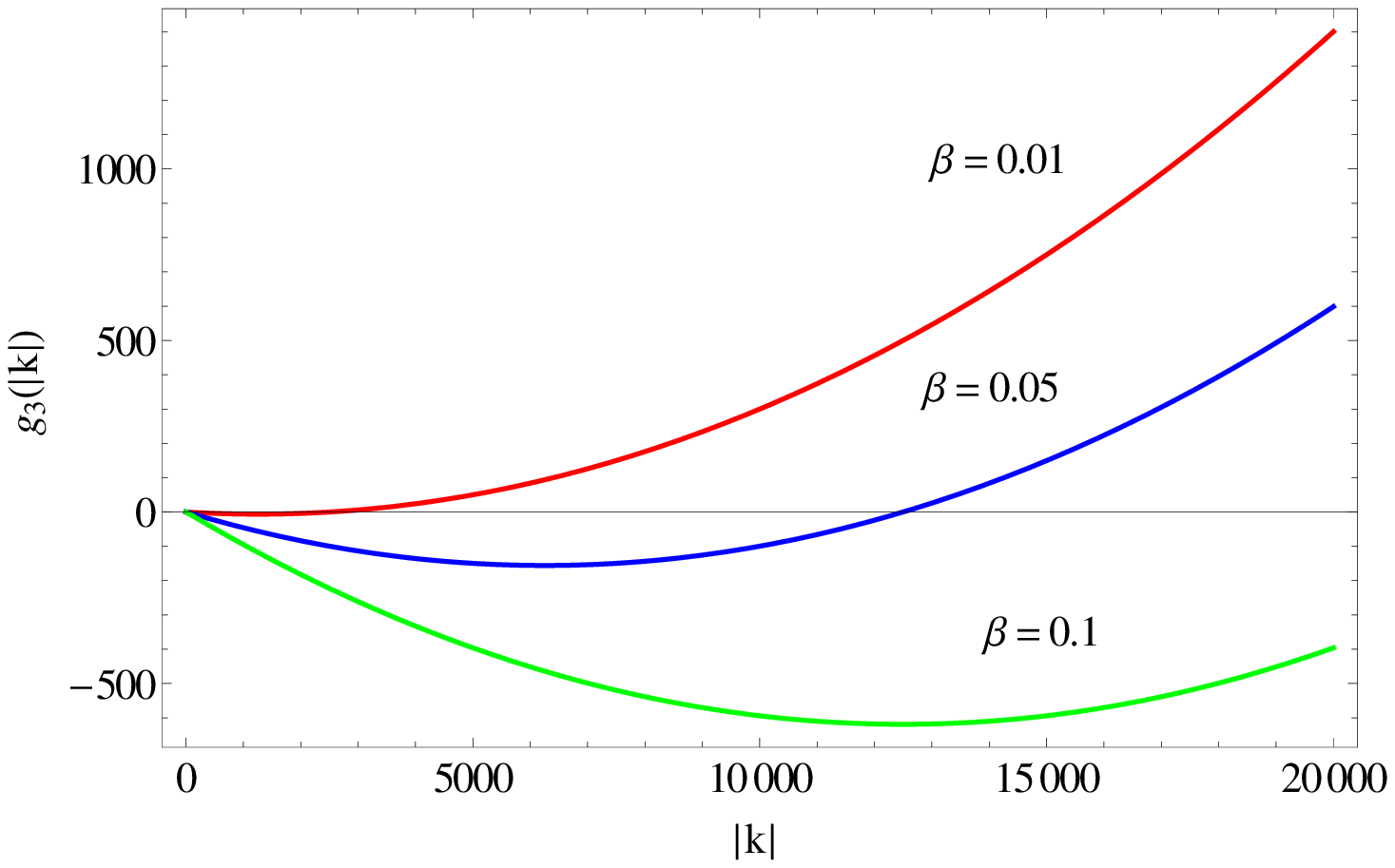}}\hspace{1em}%
{}{\includegraphics[width=0.45\textwidth]{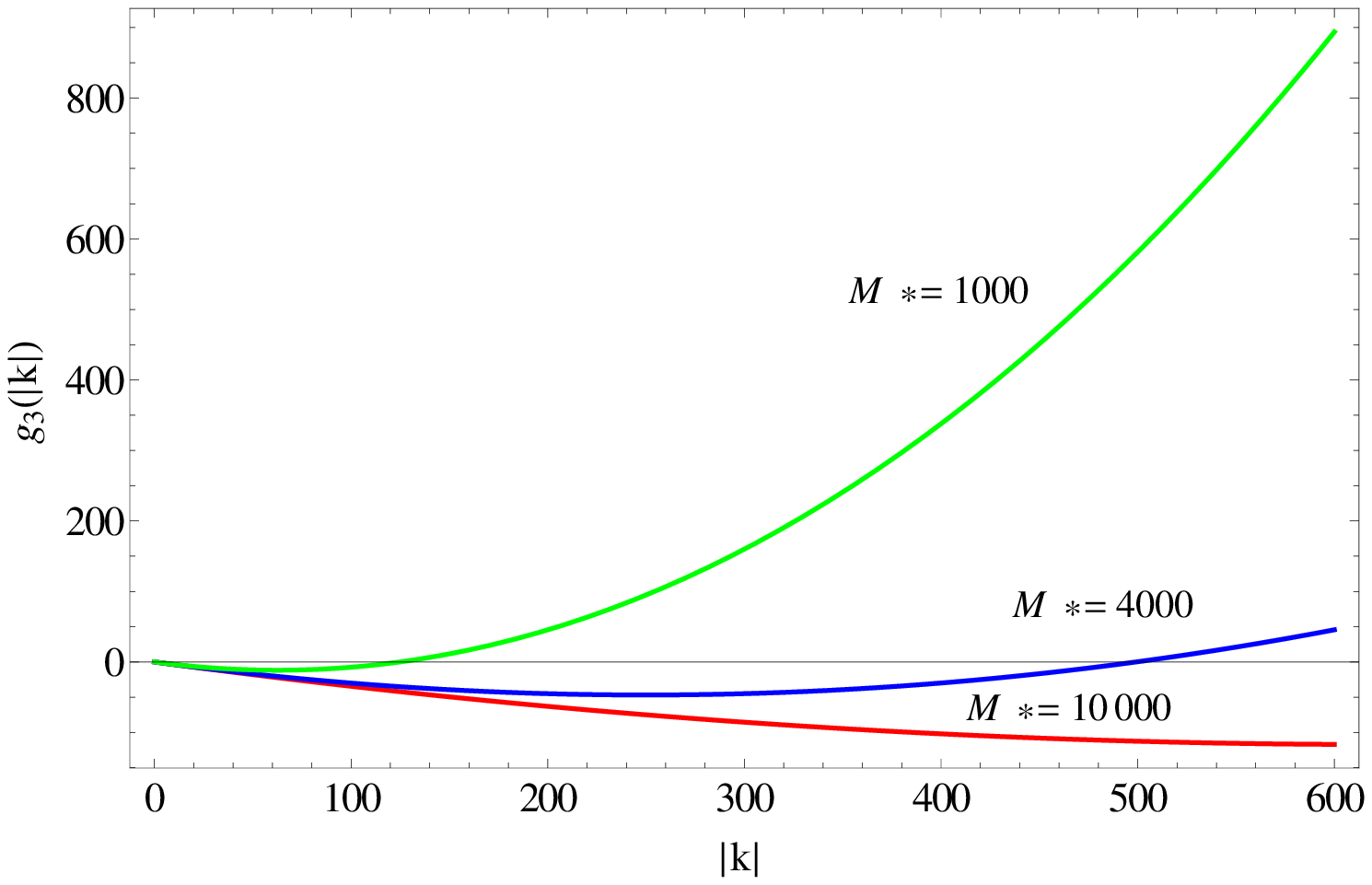}}\hspace{1em}%
\caption{Left Panel: This plot shows the variation of $g_3$ with $|\k|$ for fixed $M_*$ and varying $\beta$. We have taken $M_*=10^6$ and $\eta=1$. Right Panel: This plot shows the variation of $g_3$ with $|\k|$ for fixed  $\beta$ and varying $M_*$.}
\label{sigmabr_c1}
\end{figure}

We now turn to the question of the rate at which such transitions happen. We would like to acertain if the rate can be made lower by lowering the polymer length scale. If that were the case, it would not be possible to experimentally verify the polymer scalar field theory on the basis of this phenomenon alone. We argue that this is not the case. To solve for the rate for a given $\omega$ one must solve for $(|\k|,\theta,\phi)$ for which  
\begin{align}
\label{rate}
\omega +g_3(|\k|,\theta,\phi) =0
\end{align}

Then the integral over momentum space reduces to an integral over these values of $(|\k|,\theta,\phi)$. Let us consider the case where a single point contributes to the integral $|\k|=k_*, \theta=0, \phi=0$. This corresponds to the case when $\omega = \omega_*$. Then the momentum space integral is given by - 
\begin{align}
\label{rate2}
\nonumber
&\int \sin\theta d\theta d\phi \frac{|\k|^2 d\k}{(2\pi)^2} |c_3(\lambda_*^2 k_*)|^2\delta \left( g_3(k_*) + \omega \right)\\
&=k_*^2\frac{|c_3(\lambda_*^2 k_*)|^2}{(2\pi)^2}
\end{align} 

Now we consider the case when the polymer length scale $\lambda_*^2$ is very small compared to  $|\k|$. In this case approximate values of $c_3(\lambda_*^2 k_*)$ are available \cite{ Hossain:2010eb} (We can see from the plots that for sufficiently small values of $\beta$ $|\k|<<M_*$ so this should be a valid approximation). This is given up to first order in $g$ by:

\begin{align}
c_3 = \frac{-i}{\sqrt{2|\k|}}\left(1-\frac{3\lambda_*^2 k_*}{4} \right)
\end{align}

Substituting this back in \eqref{rate2} we obtain 
\begin{align}
\text{rate} \propto |\k|\left(1-\frac{3\lambda_*^2 k_*}{4} \right)^2
\end{align}

This shows that making $\lambda_*^2$ smaller actually increases the rate of clicking. Even when $\lambda_*$ is arbitrarily small and one would expect agreement with the usual result, the rate does not vanish. Thus our estimate suggests that the polymer scalar theory should be testable in \textit{low energy} experiments.

\section{Summary and Conclusion}

In this paper we have investigated the response of an Unruh Dewitt detector coupled to the polymer scalar field of \cite{ Hossain:2010eb}. As Lorentz invariance is lost in this theory and there is a preferred frame. We investigated three different cases : (i) The detector is at rest in the preferred frame (ii) The detector is moving with constant speed with respect to the preferred frame and (iii) the detector is moving with constant acceleration with respect to the preferred frame. 

We saw that the detector at rest does not click, while the accelerating detector does. These are true of a detector coupled to a Fock quantized scalar field as well. However we did not calculate the rate of absorption and ascertain the distribution of the energy eigenstates when the detector is at equilibrium. When the field is Fock quantized this is given by the Boltzmann distribution and the detector is therefore said to experience a thermal bath. The distribution of the energy eigenstates for the accelerated observer in the polymer case remains to be ascertained. 

The more interesting result was obtained for a detector moving at constant velocity with respect to the preferred frame. In this case we saw that there was a finite probability that the detector would click. It were in fact the (comparative) low frequency modes of the field which contributed to this radiation. It was shown that lowering the characteristic polymer length scale increases the rate of transitions.  This suggests that the polymer scalar field theory of \cite{ Hossain:2010eb} may be testable through low energy experiments that are already accessible to us. Furthermore, such experiments should put a lower limit of the polymer length scale. However we obtained only a rough estimate of this rate. A more careful numerical study needs to be done in order to make contact with experiments. 

\begin{acknowledgements}
The author would like to express his gratitude towards Ghanashyam Date and Alok Laddha for helpful discussions and comments on the draft. 
\end{acknowledgements}

(\textit{Note Added:} In a paper that appeared after our submission\cite{Husain:2015tna}, a more detailed investigation of transition rates for the inertial Unruh-DeWitt detector was undertaken, both analytically and numerically. This paper bears out our result that the inertial detector coupled to a polymer quantized scalar field can click. Furthermore, it showed that there exists critical velocity $\beta_c =1.3675$ for detectors. A detector moving below this velocity (with respect to the preferred frame) will not get spontaneously excited. However, a detector moving with a speed above $\beta_c$ will click, even when the detector's energy gap is very small or the polymer length scale $\lambda_*$is very small. In fact the rate of transitions for such a detector was shown to be proportional to $\frac{1}{\lambda_*}$. It is to be noted that the critical velocity discovered in \cite{Husain:2015tna} is well within the range of present day experiments.)


\begin{thebibliography}{20}

\bibitem{Hossain:2010eb}
  G.~M.~Hossain, V.~Husain and S.~S.~Seahra,
  Phys.\ Rev.\ D {\bf 82} (2010) 124032
  [arXiv:1007.5500 [gr-qc]].


\bibitem{Ashtekar:2002sn}
  A.~Ashtekar, S.~Fairhurst and J.~L.~Willis,
  Class.\ Quant.\ Grav.\  {\bf 20} (2003) 1031
  [gr-qc/0207106].


\bibitem{Hossain:2014fma}
  G.~M.~Hossain and G.~Sardar,
  arXiv:1411.1935 [gr-qc].


\bibitem{Rovelli:2014gva}
  C.~Rovelli,
  arXiv:1412.7827 [gr-qc].


\bibitem{Unruh:1976db}
  W.~G.~Unruh,
  Phys.\ Rev.\ D {\bf 14} (1976) 870.


\bibitem{DeWitt:2003pm}
  B.~S.~DeWitt,
  Int.\ Ser.\ Monogr.\ Phys.\  {\bf 114} (2003) 1.

\bibitem{Husain:2015tna} 
  V.~Husain and J.~Louko,
  arXiv:1508.05338 [gr-qc].

\bibitem{Ashtekar:2002vh}
  A.~Ashtekar, J.~Lewandowski and H.~Sahlmann,
  Class.\ Quant.\ Grav.\  {\bf 20} (2003) L11
  [gr-qc/0211012].


\bibitem{Kajuri:2014kva}
  N.~Kajuri,
  arXiv:1406.7400 [gr-qc].


\bibitem{G.:2013lia}
  J.~F.~Barbero G., J.~Prieto and E.~J.~S.~Villaseñor,
  Class.\ Quant.\ Grav.\  {\bf 30} (2013) 165011
  [arXiv:1305.5406 [gr-qc]].


\bibitem{Hossain:2015xqa}
  G.~Mortuza Hossain and G.~Sardar,
  Phys.\ Rev.\ D {\bf 92} (2015) 2,  024018
  [arXiv:1504.07856 [gr-qc]].

\bibitem{Satz:2006kb} 
  A.~Satz,
  Class.\ Quant.\ Grav.\  {\bf 24}, 1719 (2007)
  [gr-qc/0611067].


\bibitem{Louko:2007mu} 
  J.~Louko and A.~Satz,
  Class.\ Quant.\ Grav.\  {\bf 25}, 055012 (2008)
  [arXiv:0710.5671 [gr-qc]].
\end{thebibliography}
\end{document}